# Isoprenaline increases Excursive Restitution Slope
# in the Conscious Rabbit with Ischaemic Heart Failure

Tomofumi Kimotsuki[1], Noriko Niwa[2], Martin N. Hicks[3], Michael Dunne[3], Stuart M. Cobbe[3], Mari A. Watanabe[1,3]

[1] Internal Medicine Department, St. Louis University School of Medicine, St. Louis, USA

[2] Developmental Biology Department, Washington University School of Medicine, St. Louis, USA

[3] Cardiology Department, Glasgow Royal Infirmary, Glasgow, UK

Corresponding Author:
Mari A. Watanabe, MD, PhD
Cardiology Division
Dept. of Internal Medicine
St. Louis University Hospital
3635 Vista Avenue at Grand Avenue
St. Louis, MO 63110-0250
USA
Tel number: +1-314-577-8332
Fax number: +1-314-268-5138
watanabe@slu.edu

running title: QT dynamicity in heart failure
word count: 4600
TOC category: cardiovascular

*Abstract:*

Background: An increased QT/RR slope is hypothesized to be predictive of sudden cardiac death after myocardial infarction. Previous studies have shown that beta-adrenergic stimulation increases QT/RR slope, but the effects of beta-adrenergic stimulation on QT/RR slope in heart failure are unknown. Methods: New Zealand White rabbits underwent coronary ligation (n=15) or sham surgery (n=11), and implantation of a pediatric pacemaker lead in the right ventricle for chronic ECG recording. Eight weeks after surgery, unsedated rabbits were given intravenous administrations of 0.25~2.0 ml of 1 µmol/l isoprenaline, while peak QRS to QRS (RR) and Q to T peak (QT) intervals were measured. Results: Ligated rabbits (n=6) had lower LVEF than sham rabbits (n=7, p<.0001), but similar baseline RR (269 ± 15 *vs* 292 ± 23 ms, p=.07), QT (104 ± 17 *vs* 91 ± 9 ms, p=.1) and minimum RR (204 ±11 *vs* 208 ± 6 ms, p=.4) intervals induced by isoprenaline (0.79 ± 0.18 *vs* 0.73 ± 0.14 ml, p=.6). Hysteresis in QT *vs* TQ interval plots displayed biphasic restitution and regions of negative slope. The slope of the positive slope region was >1 in ligated rabbits (1.27 ± 0.66) and <1 in sham rabbits (0.35 ± 0.14, p=.004). Absolute value of the negative slope was greater in ligated rabbits (-0.81 ± 0.52 *vs* -0.35 ± 0.14, p=.04). Conclusion: Ischaemic heart failure produces steeper restitution slopes during beta-adrenergically induced QT/TQ hysteresis. This could underlie the propensity of failing hearts to arrhythmias.

## Introduction

The electrocardiographic QT interval plotted against heart rate, or more commonly its reciprocal the beat to beat interval, [1,2] is called QT dynamicity by clinicians [1,3]. Its *in vitro* correlate, the duration of action potential plotted against preceding diastolic interval, is called restitution curve. There is an idea that the slope of QT dynamicity or restitution curves could be used to predict sudden cardiac death, based on what is called the restitution hypothesis [4]. Computer simulations and theoretical analyses from the branch of mathematics of non-linear dynamics determined in the 1980's and early 1990's, that a steep (> 1) restitution curve slope was conducive to alternans, reentrant arrhythmias, and chaotic behavior [5-8]. An animal study showing that the slope of the restitution curve during ventricular fibrillation was greater than 1 [9] and two animal studies showing that reducing the restitution slope converted ventricular fibrillation to ventricular tachycardia [10,11], gave further support to the idea that restitution slope was an important factor in the generation of arrhythmias. There also exist human studies partially consistent with the restitution hypothesis. That is, patients with sudden cardiac death [12,13] or long QT syndrome [14] have been found to have steeper slopes of their QT dynamicity curves compared to control subjects. However, theoretical predictions have not been met in human studies in that they have failed to demonstrate QT dynamicity slope values remotely close to 1. For instance, in the Chevalier study [12] of patients 9-14 days post-myocardial infarction, median QTe/RR slope was a mere 0.13. The failure to observe a sufficiently steep slope may explain why some clinical studies find that a shallower, not steeper, QT/RR slope is related to sudden cardiac death [15,16].

The goal of the present study was to see if QT/RR slope could be pushed above the theoretically desired value of 1 by beta-adrenergic stimulation in a rabbit model of ischaemia-induced congestive heart failure. Arrhythmia incidence is increased in heart failure, so if the restitution hypothesis were correct, one would expect restitution curve slope to be increased in heart failure. Contrary to expectation, the restitution curve slope is flattened in experimentally induced congestive heart failure [17], which argues against the restitution hypothesis. However, there are studies in semi-isolated rabbit heart and man that show that sympathetic nerve stimulation and beta-adrenergic stimulation increase the slope of the restitution curve in normal hearts [18,19]. Indeed, consistent with the restitution hypothesis, increased sympathetic tone increases ventricular tachyarrhythmias [20,21] and high vagal tone protects against them [22,23]. Therefore, we hypothesized that it might be possible in the setting of transient increases in sympathetic activity to produce QT/RR slope greater than 1 in an ischaemic heart failure model, despite the presence of a lower slope in baseline (non-sympathetically stimulated) conditions.

In order to approximate spontaneously arising restitution slopes obtained from ambulatory 24 hour ECG recordings in man, we chose to study truly dynamic, as opposed to static, restitution slope in conscious animals, and named this new category of restitution, excursive restitution. We did not use pacing protocols to study restitution, because i*n vivo, s*ympathetic excitation simultaneously affects sinus rate and QT interval. Standard [24] and so-called dynamic restitution [9] slopes are both measured during pacing protocols with a fixed heart rate at its core, and are essentially static measures, despite the name of the latter. Put another way, paced protocols by nature control heart rate, and are therefore incapable of allowing measurement of restitution (the QT- heart rate relationship) during continuous heart rate change that occurs *in vivo* with sympathetic stimulation, such as during exercise and with startle.

**Methods**

*Ethical Approval*

All surgical, chronic maintenance, and data collection procedures conformed to the standards for animal care in the U.K., specifically, the Animals (Scientific Procedures) Act of 1986.

*Surgical procedure*

A well-characterized (with respect to histology, electrophysiology, and cardiac function) model of heart failure induced by chronic left ventricular infarction in the rabbit was used in this study [25-28]. Male New Zealand White rabbits, 3 – 4 kg in weight (n=26), were premedicated with a fentanyl-fluanisone combination (Hypnorm, Janssen) 0.3 ml/kg intramuscularly.  They were intubated and connected to a respirator delivering oxygen and halothane.  Using aseptic technique, the pericardium was opened and usually two but up to four ties (4/0 Ethibond) were placed around the left anterior descending coronary arteries to achieve a blanching of approximately 40% of the ventricular surface (n=15).  Quinidine was given intravenously just prior to each ligation and allowed time to wear off before the pericardial sac was closed. A pediatric defibrillator was used with the paddles in direct contact with the heart when necessary. In the remaining 11 rabbits, sham surgery was conducted in which the pericardium was opened, then sutured closed without coronary ligation.

During the same surgery, bipolar permanent pediatric pacemaker leads (Pacepath 820,

Pacesetter Systems) were placed in the right ventricle for future recording of intracardiac ECG following the technique of Manley *et al* [29]. Briefly, a pacemaker lead was introduced into the right internal jugular vein via a neck incision after excising the tines. The tip of the lead was positioned in the apex of the right ventricle using x-ray visualisation. After confirming the presence of a suitable endocardial electrocardiogram, the pacemaker lead was secured to the jugular vein by sutures. The distal end of the pacemaker lead was tunneled subcutaneously to be exteriorised via a mid-dorsal incision. At the end of the surgical procedures, the rabbit was put in an upper body jacket with zippered pockets for storing and protecting the external parts of the pacemaker leads from damage by the rabbit. Because the pacemaker lead had a ring to tip distance similar to the length of the rabbit ventricle, the recording approximated a unipolar recording of ventricular activation.

*Assessment of cardiac function*

Ultrasound was conducted on rabbits in the 7th week after surgery under sedation with Hypnorm. Standard echocardiographic measurements of cardiac dimensions were made. Left ventricular ejection fraction was computed from online planimetry measurements of the left ventricular cavity in the short axis view. Although the sonographer did not know the surgical status of the animals, the hypokinetic motion of ventricular walls if present, differentiated ligated rabbits from sham rabbits.

*Electrocardiographic recordings*

Intracardiac ECG recordings were made in conscious, unsedated animals on two separate occasions, namely in the 7th and 8th weeks after surgery. The animals were placed in a wooden box slightly larger than the animal with a hole in the front and sides allowing the animal to breathe and see outside. The external pacemaker lead was removed from the jacket and threaded through a hole in the side of the box and connected to an ECG amplifier and data acquisition system (MP100 Biopac Systems, Inc.). ECGs were recorded at a rate of 1000 Hz with the room lights dimmed and with a minimum of ambient noise. Filtering was set to pass 0.15 Hz to 35 Hz signals. Figure 1 shows ECG samples obtained from a sham and a ligated animal at baseline. After recording a baseline intracardiac ECG, a Venflon intravenous cannula (22 gauge) was inserted into one marginal ear vein and secured by adhesive tape. When the rabbit's heart rate returned to baseline, 1.0 µmol/l isoprenaline was given intravenously as a bolus in volumes ranging from 0.25 to 2.0 ml to achieve a target heart rate of 300 beats per minute. Isoprenaline stock solutions were preserved by addition of ascorbic acid and titrated to achieve a pH of 4.

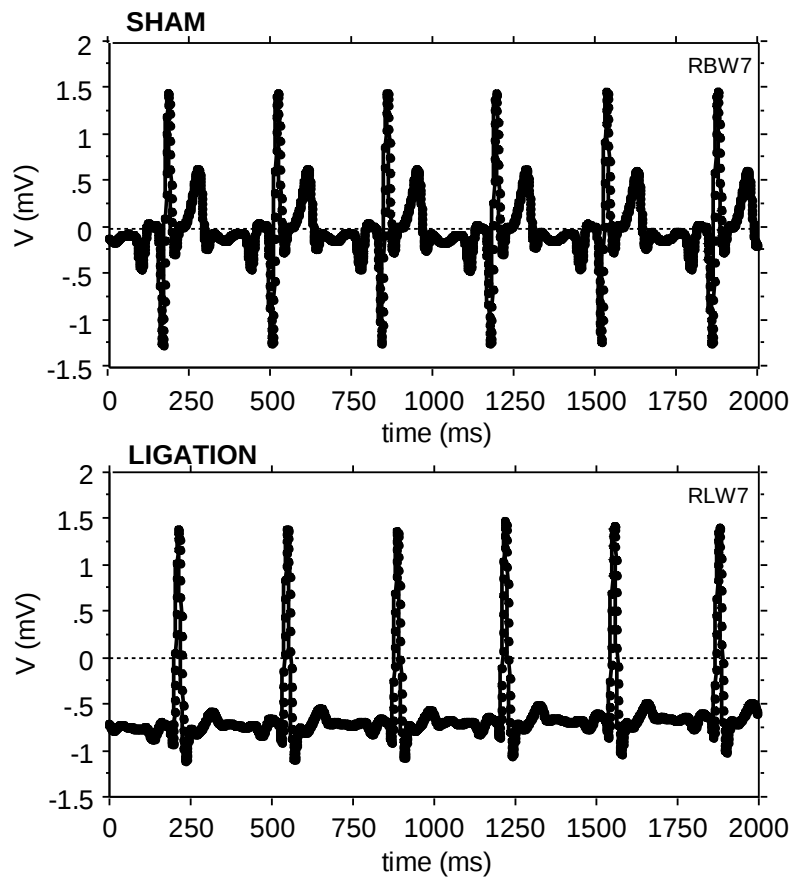

Figure 1. Intracardiac ECG examples. Top panel: sham surgery rabbit. Bottom panel: coronary ligated rabbit. These baseline recordings were made 7 weeks after surgery in the conscious state.

*Data analysis*

Offline analysis of ECG files was conducted using a program written in the Python programming language. R peaks and T peaks were found automatically and overlaid 10 beats at at time to allow manual correction of misclassified peaks. RR interval was measured as the time from R peak to R peak of the QRS waveform. QT interval was measured as the time from R peak to T peak. Ectopic and post-ectopic beats obvious on visual inspection were excluded from plots and further data analysis. Figure 2 shows examples of temporal RR and QT changes produced by 4 doses of isoprenaline from a sham and a ligated animal.

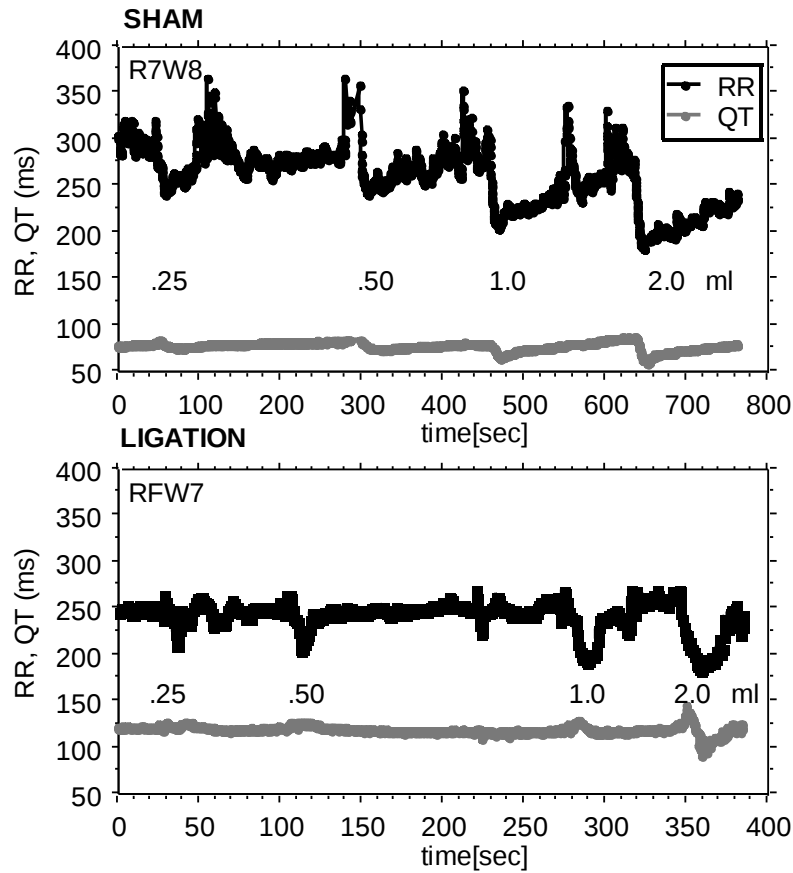

**Figure 2.** Examples of RR and QT interval change produced by different volumes of isoprenaline infusions in a sham surgery (top panel) and a coronary ligation surgery (bottom panel) rabbit.

For each interval of isoprenaline-induced heart rate increase and return to baseline, we plotted QT restitution, in our case, QT against the preceding TQ interval, for all of the beats in the interval. Figure 3 shows examples of QT restitution from a sham and a ligated rabbit. Because hysteresis was seen in the majority of restitution plots, we decided to divide the restitution plot into 4 segments which we named *biphasic, positive, negative,* and *recovery*. Slope of the 4 segments was calculated as the difference in QT divided by the difference in TQ (ms/ms). The figure legends for Figure 3 describe the criteria used to define the 5 points used to mark the start and end of the 4 segments. We also calculated time spent in each of the 4 segments, and divided RR difference and QT difference by these times to obtain rates of change of RR and of QT (ms/s). For each rabbit, we averaged the slope and rate values over all of the isoprenaline challenges without regard to volume of isoprenaline given, for several reasons. One was poor reproducibility of dose effects between the 7th and 8th weeks in the same animal; another was the difference between animals in heart rate response to identical dosages, resulting in higher volumes of isoprenaline being required to reach the target heart rate of 300 in some animals than others. We used unpaired t-tests for comparisons of sham and ligated rabbits. Statistical significance was inferred at p<.05. Mean values are reported with standard deviation values.

**Figure 3.** Representative examples of isoprenaline-induced QT/TQ restitution hysteresis from a sham surgery (left panel) and a coronary ligation surgery (right panel) rabbit. Overlaid are the 5 points used to define the 4 divisions of restitution plot hysteresis. 1: (tq, qt) at maximum RR before RR decrease due to isoprenaline, 2: (tq, QTmax), 3: (TQmin, qt), 4: (tq, QTmin), 5: (tq, qt) at whichever of RR or QT that recovered first to approximate point 1. RR or QT coordinate given in upper case means that that was the variable that determined the selection of the point, lower case coordinate means that that was the passive value of that variable for that point determined by the other variable. For example, point 3 is given as (TQmin, qt). This means that this point was selected because TQ interval reached its minimum with this beat, and the value of the QT interval did not play a role in this point's selection. The minimum value of TQ was usually reached within several beats of the minimum in RR interval.

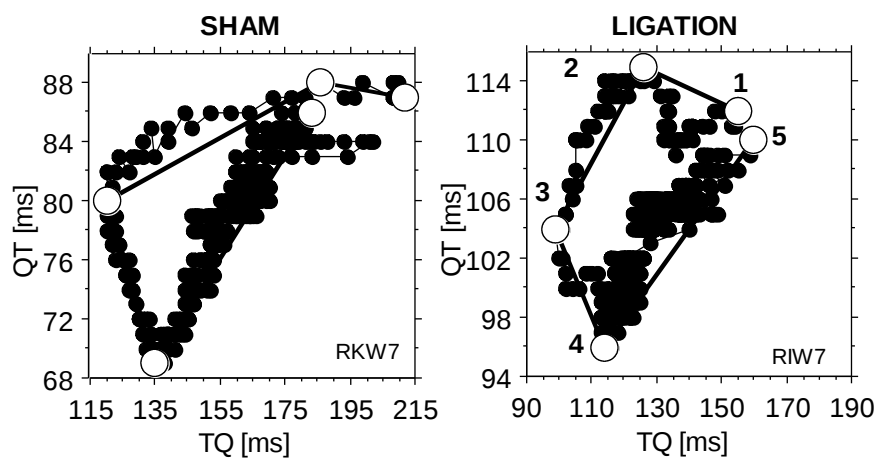

**Results**

*Study population*

None of the 11 sham surgery rabbits and five of the 15 ligation surgery rabbits died peri-operatively. Three of the perioperative deaths were caused by defibrillator malfunction. Electrocardiograms were not obtainable in 1 sham surgery and in 2 ligation surgery rabbits because of destruction of the exterior portion of the pacemaker lead by the rabbit. The T wave was not clearly identifiable in 2 sham and 2 surgery rabbits. One sham surgery rabbit was further excluded from analysis because the drops in RR value with isoprenaline challenge were > 180 ms(Figure 4). RR and QT intervals were measured in the remaining 7 sham and 6 ligation surgery rabbits. Although each rabbit was challenged with isoprenaline at least 6 times (.25 ~ 2.0 ml in two consecutive weeks), restitution slope as defined by our model of 4 segments was not analyzable approximately half of the time due to a variety of reasons. They were: lack of heart rate increase, stepped (non-monotonic) increases in heart rate, lack of QT decrease, loss of measurable T wave peak at fast heart rate due to reduction in T wave amplitude and convergence with P wave, and in one rabbit, the appearance of T wave alternans at fast heart rate. Restitution hysteresis that did not exhibit a negative slope region (coincidence of TQ minimum and QT minimum) was also excluded from analysis. The range of number of isoprenaline challenges averaged per rabbit was 2 to 5.

Table 1. Table showing similarity between sham and ligated rabbits in experimental conditions.

| | Sham (n=7) | relation | Ligation (n=6) | p |
|---|---|---|---|---|
| Ejection fraction (%) | 68 ± 2 | > | 41 ± 2 | <.0001 |
| Baseline RR (ms) | 292 ± 23 | (>) | 269 ± 15 | .07 |
| Baseline QT (ms) | 91 ± 9 | (<) | 104 ± 17 | .1 |
| Isoprenaline given (ml) | 0.73 ± 0.14 | | 0.79 ± 0.18 | .6 |
| Minimum RR achieved (ms) | 208 ± 6 | | 204 ± 11 | .4 |
| Drop in RR (ms) | 84 ± 28 | (>) | 66 ± 9 | .2 |
| Number of challenges averaged per animal | 3.4 ± 1.1 | | 3.5 ± 1.8 | .9 |

Baseline and experimental characteristics of the rabbits are given in Table 1. To summarize, left ventricular ejection fraction was significantly greater in the sham than in the ligated rabbits. Baseline heart rate had a tendency to be lower in the sham than in the ligated rabbits (p=.07). Baseline QT had a tendency to be shorter in the sham than in the ligated rabbits (p=.1). These three results were as expected. Sham and ligated rabbits did not differ in volumes of isoprenaline given (p=.6), nor in minimum RR achieved (p=.4). Figure 4 shows the drop in RR from baseline to minimum plotted against isoprenaline dose (sham n=32, ligation n=23). The tendency for isoprenaline induced RR drop to be greater for sham rabbits can be seen in this figure.

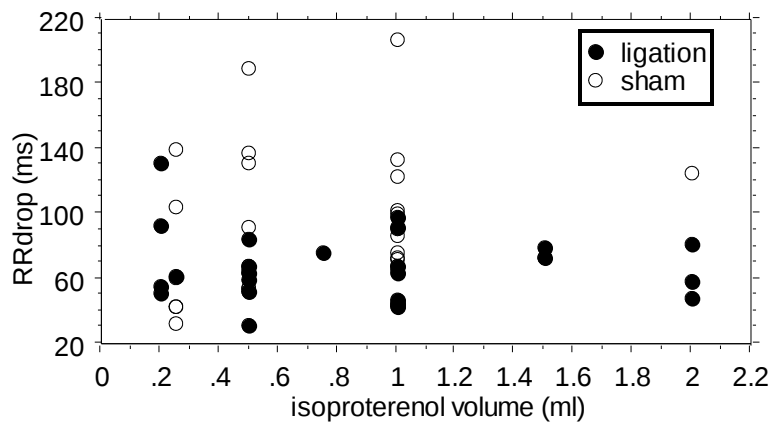

**Figure 4.** Reduction in RR value (baseline RR minus minimum RR) achieved after intravenous infusions of 1 µmol/l isoprenaline plotted against volume of infusion for 8 sham (32 isoprenaline challenges) and 6 ligated (23 isoprenaline challenges) rabbits. The two points representing RR drops > 180 were from the same rabbit. This rabbit was excluded from further analysis because inclusion | exclusion produced a statistically significant | insignificant difference in RR drop between ligated and sham rabbits (p=.025 | p=.075).

*Restitution slopes and rates of change*

Table 2 gives the QT/TQ restitution slopes and the rates of change of RR and QT intervals for the 4 hysteresis segments. Figure 5 shows box plots of these three quotients for the biphasic, positive and negative segments. We describe the segments in turn.

*Biphasic hump:* For the initial segment where QT increased although RR had begun to decrease, there was no statistically significant difference between sham and ligated rabbits in restitution slope, rate of RR drop, or rate of QT increase.

*Positive slope:* For the classically defined restitution segment where both RR and QT decreased, restitution slope was significantly greater for ligated rabbits, with an average slope value of $1.3 \pm 0.7$ for ligated, $0.4 \pm 0.1$ for sham rabbits (p=.0040). The rate of RR decrease was similar (p=.5), but QT drop rate was significantly faster (-3 ms/s) in ligated rabbits than in sham rabbits (-2 ms/s, p=.040).

*Negative slope:* For the restitution segment where QT continued to decrease although RR had begun to recover, restitution slope was significantly greater for ligated rabbits, with an average slope value of $-0.8 \pm 0.5$ for ligated *vs* $-0.3 \pm 0.1$ for sham rabbits (p=.04). The rate of RR increase was similar (p=.8), but QT drop rate was significantly faster (-3 ms/s) in ligated rabbits than in sham rabbits (-1 ms/s, p=.02). A paired comparison of QT drop rate for the ligated rabbits between the positive and negative slope segments produced a p of .47, and for the sham rabbits, .02. This suggested that in ligated rabbits, QT continued to drop at a similar rate after the TQ minimum had been reached, while in sham rabbits, the rate of QT reduction per time tapered off significantly, once the TQ minimum had been reached.

*Recovery slope:* For the restitution segment where both RR and QT increased in tandem, no significant differences were seen in any of the parameters calculated. The restitution slope averages were 0.6 for ligated and 0.3 for sham rabbits.

Time spent in each of the 4 segments was statistically similar for sham and ligated rabbits, as was change in RR, suggesting that the difference in magnitude or rate of RR response to isoprenaline could not account for the difference in restitution slope and rate of QT change between sham and ligated rabbits.

Table 2. Comparison between sham and ligated rabbits for restitution slope and rates of RR and QT interval change.

| | *Sham(n=7)* | *Relation* | *Ligation (n=6)* | *p* |
|---|---|---|---|---|
| Biphasic hump | | | | |
| dQT/dTQ | -.09 ± .09 | | -.17 ± .10 | .2 |
| dRR/dt (ms/s) | -9.7 ± 4.6 | | -6.7 ± 1.5 | .2 |
| dQT/dt (ms/s) | .58 ± .59 | (<) | 1.3 ± .68 | .07 |
| dt (s) | 5.4 ± 2.0 | | 6.6 ± 2.0 | .3 |
| Positive | | | | |
| dQT/dTQ | .35 ± .14 | < | 1.27 ± .66 | .004 |
| dRR/dt (ms/s) | -8.0 ± 3.4 | | -6.9 ± 1.1 | .5 |
| dQT/dt (ms/s) | -1.9 ± .80 | < of absolute value | -3.0 ± .87 | .04 |
| dt (s) | 5.8 ± 2.0 | | 4.5 ± .8 | .2 |
| Negative | | | | |
| dQT/dTQ | -.35 ± .14 | < of absolute value | -.81 ± .52 | .04 |
| dRR/dt (ms/s) | 2.8 ± 1.2 | | 2.4 ± 3.6 | .8 |
| dQT/dt (ms/s) | -.95 ± .27 | < of absolute value | -2.64 ± 1.62 | .02 |
| dt (s) | 6.5 ± 2.3 | | 6.9 ± 3.1 | .8 |
| Recovery | | | | |
| dQT/dTQ | .25 ± .16 | | .61 ± .97 | .4 |
| dRR/dt (ms/s) | 2.2 ± 1.9 | | 2.8 ± 2.3 | .7 |
| dQT/dt (ms/s) | .24 ± .17 | | .77 ± .82 | .1 |
| dt (s) | 67 ± 38 | | 37 ± 23 | .1 |

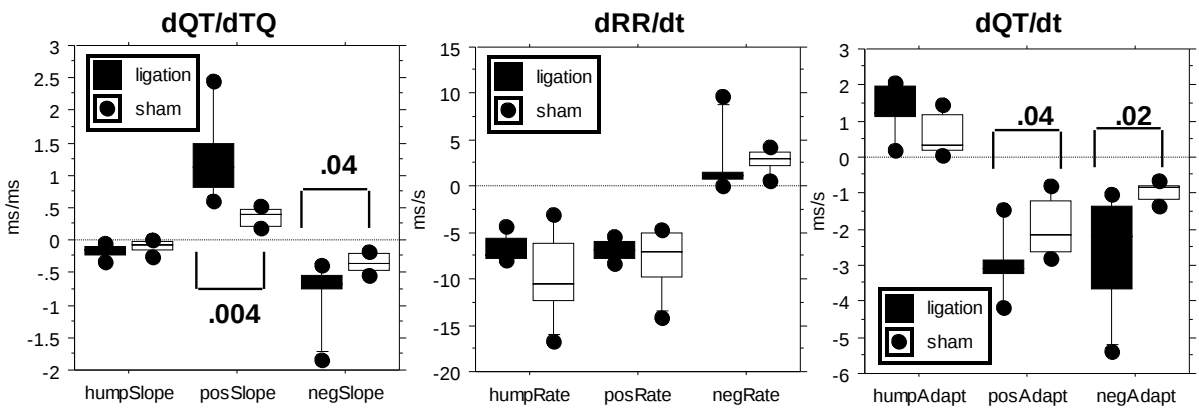

**Figure 5.** Box plots of restitution slopes, and rates of RR and QT interval change per unit time in seconds for 6 ligation surgery and 7 sham surgery rabbits. Within each panel, from left to right are the distribution of values for the biphasic hump segment, positive slope segment, and negative slope segment of QT hysteresis. The box edges represent 25th and 75th percentile, the horizontal line within the box, the median, the whiskers represent 10th and 90th percentile, and the discrete points, all of the values outside the 10th to 90th percentile. See Table 2 for mean and standard deviation values of these parameters.

**Discussion**

As stated in the introduction, the goal of the present study was to see if restitution slope could be pushed above 1 by beta-adrenergic stimulation in a rabbit model of ischaemia-induced congestive heart failure. We found that isoprenaline infusion indeed produced an average positive segment restitution slope greater than 1 (1.27) in heart failure rabbits, and a slope less than 1 (0.35) in sham operated rabbits. In addition, we observed a clear region of negative slope in many of the isoprenaline challenges, that led to hysteresis in the restitution plot. The slope of the negative segment was also steeper in heart failure rabbits. The two restitution slopes were steeper in the heart failure rabbits despite the tendency for failing rabbits to start with a faster baseline heart rate, leading them to undergo smaller increases of heart rate to the 300 b.p.m. target heart rate. Because failing and sham surgery rabbits displayed similar rates of RR change over those hysteresis segments, and because the two groups received similar doses of isoprenaline, we were also able to exclude those factors as causes for the difference in restitution slopes. In contrast, rate of change of QT per unit time was higher in the heart failure rabbits for the positive and negative slope segments, and also for the biphasic segment albeit with a p value of 0.07, suggesting that greater magnitude of QT dynamics itself was a core feature of failing hearts, along with steeper restitution.

*A new category of restitution*

The term restitution, has of late, come to be used to describe any relationship between heart rate and some measure of repolarisation, such as action potential duration, refractory period, or electrocardiographic QT interval. Heart rate is usually quantified by its reciprocal the beat to beat interval, *i.e.*, in units of time, rather than beats per minute. Beyond the choice of particular independent and dependent variables, there are also different categories of restitution, depending on the stimulation protocol used to vary the heart rate. Originally, restitution referred solely to the relationship between a single beat change in heart rate and repolarisation. Such restitution curves were obtained by stimulating cardiac tissue at a constant heart rate, followed by one perturbatory stimulus at a different heart rate, recording the repolarisation measure of the perturbed beat, then repeating the constant rate-single perturbed rate sequence with varied timing of the perturbatory stimulus [30-32]. Such a restitution curve is not unique. Depending on the constant heart rate used before the perturbatory stimulus, different restitution curves arise. To prevent confusion caused by the expansion in the meaning of the term restitution, it has been suggested that this type of classical restitution be renamed "standard restitution" [24]. Ventricular premature contractions occurring spontaneously in otherwise stable heart rate conditions in man can produce a standard restitution function, provided the coupling interval of the premature beat varies enough to provide a spread of points.

A second category of restitution refers to the relationship between constant heart rate and repolarisation. This type of restitution is obtained by stimulating cardiac tissue at a constant heart rate until the repolarisation measure stabilizes, then repeating this for other heart rates [33]. The plot of this relationship was originally called the steady state action potential duration curve [32]. In practice, investigators do not wait for the repolarisation measure to stabilize completely, but deliver a fixed number of pacing stimuli (usually 8~20), then move on to the next heart rate. Koller *et al* called the heart rate - repolarisation relationship obtained in this manner, dynamic restitution [9], and thereby expanded the meaning of the term restitution. The closest analog of dynamic restitution in man would be a plot of QT intervals compiled from different stable heart rates, such as found over the day. The equation used to derive Bazett's formula, QT= 0.44 x square root of RR, could be regarded as the dynamic restitution curve for a population [34].

In the current study, we studied a third category of restitution, one that we suggest calling excursive restitution. It tracks the heart rate - repolarisation relationship for all beats during a perturbation away from and back to baseline (steady state). During excursive restitution, there is no steady state, and the space of families of restitution curves as studied by other stimulation protocols [35,36] is traversed during the excursion. Excursive restitution happens spontaneously throughout the day in man as heart rate varies to accommodate the body's metabolic and postural needs. Studies of QT and heart rate changes induced by exercise in man also belong to this category of restitution. In contrast to standard and dynamic restitution, whose concepts were defined in animal studies, the only previous study of excursive restitution that we know of is one by Lux & Ershler who demonstrated the effects of applying a linear ramp pacing protocol to a repolarisation parameter in the dog [37].

*Negative slope in restitution*

Hysteresis in restitution is a well known phenomenon [38]. Its presence complicates the definition and measurement of restitution slope, and several different methods have been developed for circumnavigating this problem (Neilson 2000; Lang *et al*, 2001; Pueyo *et al*, 2003; Fossa 2008). We believe that our current report marks the first time the existence of negative restitution has been reported *in vivo*, for any category of restitution. Negative restitution simply represents continued QT decrease as heart rate begins its return to slower rate, i.e., it is nothing but QT response lagging behind that of RR. A schematic diagram of action potential duration transitions between two fixed heart rates from an older study suggests graphically how negative restitution could occur (Figure 3 of ref. [35]). In our opinion, demonstration of the existence of negative restitution is important because it guarantees existence of a point of infinite magnitude slope between the positive and negative segments. Although a slope greater than 1 implies instability and arrhythmia susceptibility in standard and dynamic restitution curves, the theoretical implications of having a slope of infinity in excursive restitution curves are entirely unknown at this time, and is a subject for future research.

*Effects of adrenergic stimulation on repolarisation*

Isoprenaline has been used frequently over the years to mimic sympathetic activity. Because computerized measurement techniques did not exist, early studies focused largely on measurements of discrete timepoints after isoprenaline infusion. Nevertheless, hysteresis was noted (RR decrease before QT) [43], as was dose-dependent reduction in action potential duration in fixed atrial pacing [44], and presence of biphasic change in action potential duration (i.e., an increase preceding the decrease)[45]. Sympathetic nerve stimulation was shown to shorten refractory period in epicardial and endocardial tissue by similar amounts [46] and produce biphasic change in monophasic action potential duration despite fixed atrial rate [47]. More recent studies indicate emerging interest in the restitution slope. Taggart *et al* [48] reported a value of 1.05 for the standard restitution slope using programmed stimulation in man and demonstrated that it increased with isoprenaline infusion. Ng *et al* [49] reported that sympathetic nerve stimulation led to reduced refractory period, increased restitution slope, and reduced ventricular fibrillation threshold. They further found a significant association between restitution slope and ventricular fibrillation threshold. With regard to ionic mechanism, isoprenaline's action on repolarisation has been shown to depend on the slowly activating component of the delayed rectifier potassium current (IKs). Isoprenaline enhances IKs in a dose-dependent fashion, negatively shifts IKs activation voltage dependence, and accelerates IKs activation [50]. Consistent with this result, shortening of repolarisation by isoprenaline is prevented by IKs  block *in vitro* and *in vivo* [51].

*Limitations*

We were forced to choose the T peak as our measure of repolarisation rather than the time from R peak to end of the T wave, because at fast heart rates, the T wave end became impossible to distinguish from the start of the following negative P wave. Even using the T peak, we were unable to analyze data at high heart rates in some isoprenaline challenges. A study in man reporting that the slope of daytime Q-Tpeak versus RR did not predict sudden death after myocardial infarction in contrast to Q-Tend [12] would make Q-Tend preferable for study. However, from a restitution hypothesis viewpoint, there is no reason to believe that QT dynamicity calculated using T end is superior to T peak in predicting arrhythmia vulnerability, since both reflect action potential duration in some section of the heart according to recent studies. In one *in vivo* study in swine utilizing recordings from multiple endocardial and epicardial sites, it was demonstrated that the peak and end of the T wave coincided respectively with the times of earliest and latest end of repolarisation over the whole heart [52]. In another study, the T peak and T end were reported to correspond to end of repolarisation in epicardium and M cells, respectively [53]. In a previous study using New Zealand White rabbits, the same breed used in the current study, the stimulus to T peak time recorded by the intracardiac pacemaker lead and action potential duration measured from transmembrane recordings in 6 ventricular septa were related by the regression equation, stimulus_to_Tpeak = 0.9 x APD[100%] + 25 (ms), r=0.96, over an APD range of 200 ~ 450 ms [29]. Finally, work by Nearing & Verrier [54] suggests that T wave changes such as T wave alternans are minimal in the terminal portion, and that the greatest amplitude differences reside in the segment preceding the peak (see Figure 1 of the reference). For these various reasons, we feel there is justification in using QTpeak results. Furthermore, our results may have applications to studies of repolarisation dynamics in rodents whose high heart rates make T wave end hard to identify.

The recovery segment results could not be determined as accurately as the first three segments. This was because recovery was gradual compared to the other three segments and was less well defined. In some cases, QT or RR did not recover to baseline values within the time of observation, or began to decrease while the other was still increasing. When selecting the beat defining the end of the recovery segment, we generally chose either a RR or QT value close to baseline for which the recovery segment would approximately lie over the data points in three plots, the RR *vs* time, QT *vs* time, and restitution plots. Average restitution slope value for the recovery segment was 0.6 and 0.3 in the failing and sham surgery rabbits respectively (p=.1). The difference may have reached statistical significance if the recovery segment had been defined some other way.

We were unable to assess the contribution of isoprenaline-induced increased contractility to our results. However, when we recorded the ECG, there was endocardial fibrosis around the pacemaker lead embedding the lead tip in the right ventricular apex, so that the anatomical location of the lead tip remained fixed relative to the surrounding myocardium. This would have mitigated some of the motion artifact produced by increased contractility.

A fourth limitation of our study was that many isoproterenol challenges were excluded from the final analysis for the six cases enumerated in the Results section. We discuss three of them here in more detail. i) We do not know why stepped increases in heart rate occurred. Perhaps the animals were startled by the pharmacologically induced sudden increase in heart rate and contributed their own sympathetic surge. In these cases, there were multiple measurable restitution slopes, but it was unclear how to treat them statistically. ii) In cases of lack of QT decrease, we considered treating them as zero valued restitution slope, but then the question arose of whether it made sense to give values of zero to all 4 slopes when they were ill-defined in the case of lack of QT change. E.g., recovery slope defined as recovery of both RR and QT would have to be redefined as recovery of RR only. On the other hand, there is precedence in analyzing only measurable responses in human data. Spontaneous baroreflex sensitivity is calculated as the slope between blood pressure and heart rate only for time segments over the 24 hour record in which both are increasing. Baroreflex sensitivity measured in this way correlates closely with measurements using pharmacological challenges [55], and suggests one way of dealing with the unpredictability of cardiac responses displayed *in vivo*. iii) Loss of measurable T wave peak at fast heart rate was the most frustrating, because there was a clear response in heart rate and QT, and a technique for quantifying repolarisation other than intracardiac ECG might have allowed us to follow their values to their minimum. Exclusions due to this reason happened at higher doses of isoprenaline, and might have been mitigated by increasing the number of challenges using lower doses, if we had had real time QT measurement capabilities.

**Conclusion**

Rate of change of QT per unit time and QT restitution slope during beta-adrenergic stimulation are both greater in heart failure rabbits than in control rabbits. The fact that QT restitution slope was greater than 1 in the heart failure rabbits is compatible with the increased propensity for arrhythmic death in patients with heart failure, according to the restitution hypothesis.

**Author contributions**

All authors contributed to the study: TK and NN in the analysis and interpretation of data, MNH and MD in the conception and design of the study and in conducting the experiments, SMC in the conception and design of the study, and MAW in all of the above and in writing the computer program for data analysis. All authors also read the manuscript critically and approved the final version, with the exception of MNH, who passed away before the study was completed.

**Grants, Acknowledgments, and Disclosures**

The experimental portion of this research project was conducted at Glasgow Royal Infirmary, with the financial support of British Heart Foundation Project Grant PG/02/155 and Stuart M. Cobbe. The analytical portion of this research project was conducted at St. Louis University School of Medicine. TK received salary support from Sankyo Co, Ltd., Shizuoka, Japan. The authors have no conflicts of interest to disclose.

**Figure Legends**

Figure 1. Intracardiac ECG examples. Top panel: sham surgery rabbit. Bottom panel: coronary ligated rabbit. These baseline recordings were made 7 weeks after surgery in the conscious state.

Figure 2. Examples of RR and QT interval change produced by different volumes of isoprenaline infusions in a sham surgery (top panel) and a coronary ligation surgery (bottom panel) rabbit.

Figure 3. Representative examples of isoprenaline-induced QT/TQ restitution hysteresis from a sham surgery (left panel) and a coronary ligation surgery (right panel) rabbit. Overlaid are the 5 points used to define the 4 divisions of restitution plot hysteresis. 1: (tq, qt) at maximum RR before RR decrease due to isoprenaline, 2: (tq, QTmax), 3: (TQmin, qt), 4: (tq, QTmin), 5: (tq, qt) at whichever of RR or QT that recovered first to approximate point 1. RR or QT coordinate given in upper case means that that was the variable that determined the selection of the point, lower case coordinate means that that was the passive value of that variable for that point determined by the other variable. For example, point 3 is given as (TQmin, qt). This means that this point was selected because TQ interval reached its minimum with this beat, and the value of the QT interval did not play a role in this point's selection. The minimum value of TQ was usually reached within several beats of the minimum in RR interval.

Figure 4. Reduction in RR value (baseline RR minus minimum RR) achieved after intravenous infusions of 1 μmol/l isoprenaline plotted against volume of infusion for 8 sham (32 isoprenaline challenges) and 6 ligated (23 isoprenaline challenges) rabbits. The two points representing RR drops > 180 were from the same rabbit. This rabbit was excluded from further analysis because inclusion | exclusion produced a statistically significant | insignificant difference in RR drop between ligated and sham rabbits (p=.025 | p=.075).

Figure 5. Box plots of restitution slopes, and rates of RR and QT interval change per unit time in seconds for 6 ligation surgery and 7 sham surgery rabbits. Within each panel, from left to right are the distribution of values for the biphasic hump segment, positive slope segment, and negative slope segment of QT hysteresis. The box edges represent 25th and 75th percentile, the horizontal line within the box, the median, the whiskers represent 10th and 90th percentile, and the discrete points, all of the values outside the 10th to 90th percentile. See Table 2 for mean and standard deviation values of these parameters.